\documentclass[prb,preprint,showpacs]{revtex4}%
\usepackage{amsfonts}
\usepackage{amsmath}
\usepackage{amssymb}
\usepackage{multirow}
\usepackage{graphicx}%
\providecommand{\U}[1]{\protect\rule{.1in}{.1in}}
\providecommand{\U}[1]{\protect\rule{.1in}{.1in}}
\providecommand{\U}[1]{\protect\rule{.1in}{.1in}}
\providecommand{\U}[1]{\protect\rule{.1in}{.1in}}
\providecommand{\U}[1]{\protect\rule{.1in}{.1in}}
\providecommand{\U}[1]{\protect\rule{.1in}{.1in}}

\begin{document}

\title{Dissociation of hydrogen molecules on the
clean and hydrogen-preadsorbed Be(0001) surface}

\author
{Yanfang Li$^{1,2}$, Yu Yang$^2$, Bo Sun$^2$, Yinghui Wei$^1$, Ping
Zhang$^{2,3}$\footnote{Corresponding author. E-mail address:
zhang\underline{ }ping@iapcm.ac.cn}}

\affiliation{$^1$College of Materials Science and Engineering,
Taiyuan University of Technology, Taiyuan 030024, People's Republic
of China}%
\affiliation{$^2$LCP, Institute of Applied Physics and Computational
Mathematics, P.O. Box 8009, Beijing 100088, People's Republic of
China}%
\affiliation{$^3$Center for Applied Physics and Technology, Peking
University, Beijing 100871, People¡¯s Republic of China}%

\date{\today}

\pacs{68.43.Bc, 68.43.Fg, 68.43.Jk, 73.20.Hb}

\begin{abstract}
Using first-principles calculations, we systematically study the
potential energy surfaces and dissociation processes for hydrogen
molecules on the clean and hydrogen-preadsorbed Be(0001) surfaces.
It is found that the most energetically favored dissociation channel
for H$_2$ molecules on the clean Be surface is at the surface top
site, with the minimum energy barrier of 0.75 eV. It is further
found that after dissociation, hydrogen atoms do not like to cluster
with each other, as well as to penetrate into subsurface sites. For
the hydrogen-preadsorbed Be(0001) surface, the smallest dissociation
energy barrier for H$_2$ molecules is found to be 0.50 eV, which is
smaller than the dissociation energy barrier on a clean Be(0001)
surface. The critical dependence of the dissociation energy barriers
for H$_2$ molecules on their horizontal distances from the
preadsorbed hydrogen atom is revealed. Our studies well describe the
adsorption behaviors of hydrogen on the Be(0001) surface.

\end{abstract}

\maketitle

\section{Introduction}

The dissociative adsorption of hydrogen molecules on metal surfaces
is of great importance since it is a crucial step in many
application processes, such as hydrogen storage for fuels
\cite{Hu2002}, hydrogen caused embrittlement \cite{Lu2005}, and
heterogeneous catalysis \cite{King1988,Gustafsson2006}. Among all
the metals, simple $sp$-metals (metals with only $sp$ valence
electrons), such as Be and Mg, are good candidates for hydrogen
storage materials, and have vast applications in industries where
hydrogen caused embrittlement need to be prevented. So the
adsorption and dissociation of hydrogen molecules on these
$sp$-metals need special concerns. For the H$_2$/Mg(0001) system,
many studies have already been carried out
\cite{Vegge2004,Johansson2006,Wu2008}. It is found that the most
energetically favorable site for the dissociation of molecule H$_2$
on the Mg(0001) surface is the bridge site \cite{Vegge2004,Wu2008},
and the theoretically calculated energy barrier for dissociation is
in good accordance with the experimentally observed values
\cite{Wu2008}. Similar to the Mg(0001) surface, the Be(0001) surface
also has considerable $s$ and $p$ electronic states distributing
around the Fermi energy \cite{Chulkov1987}, which makes it a good
prototype to study the adsorption behaviors of H$_2$ on $sp$-metals.
However, theoretical studies have not been carried out on the
dissociation process of H$_2$ molecules on the Be(0001) surface yet.

On the other hand, Be has vast applications in modern nuclear
devices. In the international thermonuclear experimental reactor
(ITER), Be contributes the major part of the first wall
\cite{Allouche08}, based on its low atomic number, ability to remove
oxygen from the plasma, and ability to adsorb residue gases composed
of light atoms as C, H and O \cite{Causey02,Argentina2000,
Zalkind2002}. For these important applications, the adsorption of
small molecules such as H$_2$, O$_2$ and H$_2$O on the Be surfaces
need to be specially investigated. Actually, the molecular
adsorption and dissociation of H$_2$O \cite{Zalkind1997,Gomez07} and
O$_2$ \cite{Linsmeier00,Zalkind05,Zhang09} have already been studied
systematically. Therefore, studying the adsorption of hydrogen on
the Be(0001) surface not only gives us knowledge about the common
characters of interactions between H$_2$ and $sp$-metals, but also
provides us information on how to advance the usages of Be materials
in nuclear reactors.

Correspondingly, the adsorption of hydrogen atoms on the Be(0001)
surface has been widely studied, including the adsorption of
different coverage \cite{Feibelman1993,Stumpf1995,Pohl04}. It is
found that at low coverage, H atoms choose to adsorb at surface hcp
hollow sites, while at the 1 monolayer (ML) coverage, the adsorption
at surface bridge sites is energetically more favorable
\cite{Stumpf1995}. However, as the first step towards surface
hydrogenation, the adsorption and dissociation of molecular H$_2$ on
the Be(0001) surface is more important for the above research
backgrounds. And to our surprise, in contrary to the vast studies on
the adsorption of atomic H, the adsorption and dissociation of
molecular H$_2$ on the Be(0001) surface has not been studied at all.
Based on these backgrounds, here we perform first-principles
calculations to systematically study the adsorption and dissociation
of molecular H$_2$, the distribution of dissociated H atoms, and the
adsorption and dissociation of molecular H$_2$ on the
hydrogen-preadsorbed Be(0001) surfaces. The rest of this paper is
organized as follows. In Sec. II, we give details of the
first-principles total energy calculations, which is followed in
Sec. III by our results for the PESs of molecular H$_2$ on the clean
Be(0001) surface, where the energetically favored dissociation paths
and the dissociation energy barriers are presented and discussed. In
Sec. IV, we study the diffusion and penetration of hydrogen atoms on
the Be(0001) surface, and discuss the distribution properties of the
dissociated hydrogen atoms. Then the PESs for molecular H$_2$ on
hydrogen-preadsorbed Be(0001) surface are calculated and discussed
in Sec. V, similar to what we did in Sec. III. And finally in Sec.
VI, we give our conclusions.

\section{Methods}

Our calculations were performed within density functional theory
(DFT) using the Vienna {\it ab-initio} simulation package (VASP)
\cite{VASP}. The PW91 \cite{PW91} generalized gradient approximation
and the projector-augmented wave potential \cite{PAW} were employed
to describe the exchange-correlation energy and the electron-ion
interaction, respectively. The cutoff energy for the plane wave
expansion was set to 300 eV. The Be(0001) surface was modeled by a
slab composing of five atomic layers and a vacuum region of 20 \AA.
The $2\times2$ and $3\times3$ supercell (in which each monolayer
contains four and nine Be atoms) were adopted respectively in the
study of the H$_2$ adsorption on the clean and hydrogen-preadsorbed
Be(0001) surfaces. Our test calculations have shown that the
$2\times2$ supercell is sufficiently large to avoid the interaction
between adjacent hydrogen molecules. And the $3\times3$ supercell
was used to explore more carefully the effect of the preadsorbed H
atom. Integration over the Brillouin zone was done using the
Monkhorst-Pack scheme \cite{Monkhorst} with $11\times11\times1$ and
$7\times7\times1$ grid points respectively for the $2\times2$ and
$3\times3$ supercell. A Fermi broadening \cite{Weinert1992} of 0.1
eV was chosen to smear the occupation of the bands around the Fermi
energy (E$_{F}$) by a finite-$T$ Fermi function and extrapolating to
$T=0$ K. The calculation of the potential energy surface for
molecular H$_2$ was interpolated to 209 points with different bond
length ($d_{\rm H-H}$) and height ($h_{\rm H_2}$) of H$_{2}$ at each
surface site. The calculated lattice constant of bulk Be ($a$, $c$)
and the bond length of a free H$_{2}$ molecule are 2.26 \AA, 3.56
\AA~ and 0.75 \AA, respectively, in good agreement with the
experimental values of 2.285 \AA, 3.585 \AA~\cite{Wachowicz2001} and
0.74 \AA~\ \cite{Huber1979}.

\section{Adsorption and dissociation of molecular H$_2$ on the clean Be(0001) surface}

The geometry and electronic properties of the clean Be(0001) surface
is first investigated. In comparison with that in bulk Be, the
electronic density of states (DOS) at the Fermi energy (E$_F$) has a
relatively small value, the DOS at E$_{F}$ in the clean Be(0001)
surface is prominently enhanced \cite{Zhang09}. From wavefunction
analysis, it is found that the electronic states around E$_{F}$  are
mainly Be $2p$ states, and mainly accumulate within the two topmost
Be layers. Due to this pronounced surface charge redistribution, the
two outmost Be(0001) layers relax significantly from the bulk
values. The first-second interlayer contraction is 3.8\% and the
second-third interlayer expansion is nearly 1.2\%, which is in
agreement with recent first-principles calculations
\cite{Lazzeri1998} and comparable with experimental measurements
\cite{Pohl1998}.

\begin{figure}
\includegraphics[width=0.8\textwidth]{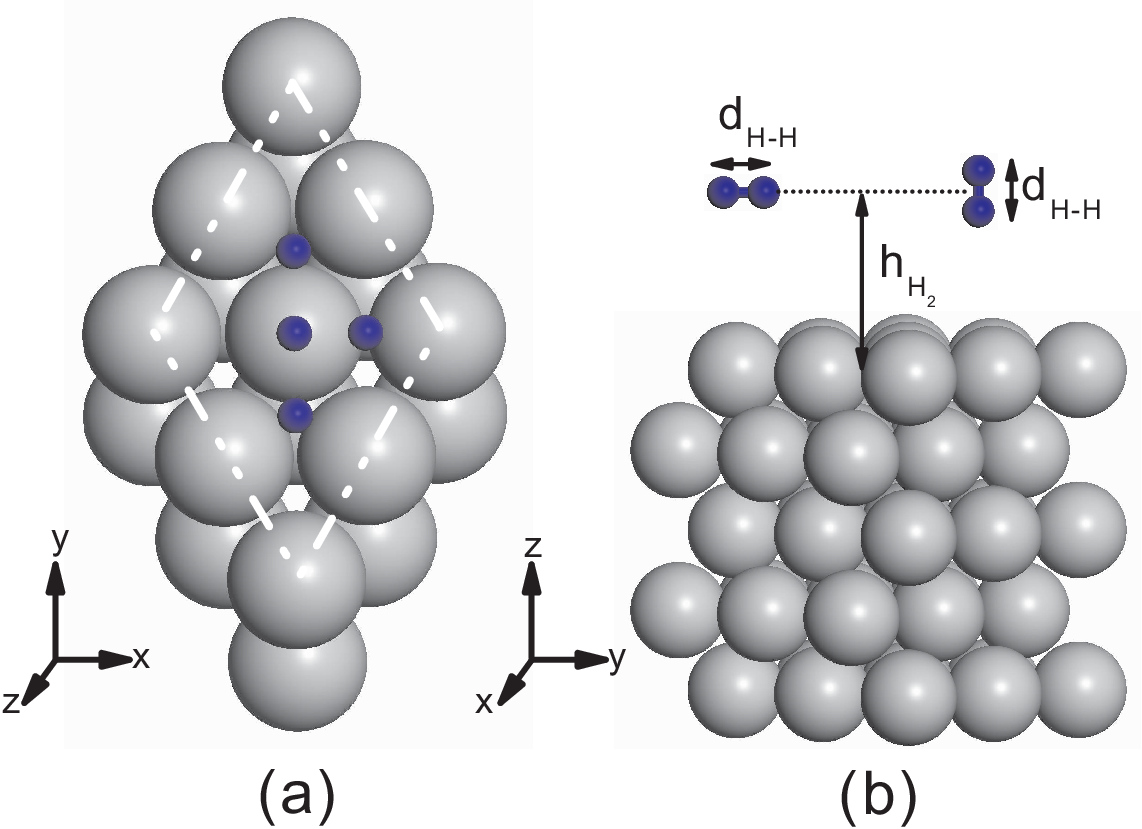}
\caption{(Color online). (a) The $p$($2\times2$) surface cell of
Be(0001) and four on-surface adsorption sites. Here only the outmost
two layers of the surface are shown. (b) The sketch map showing that
the molecule (with vertical or parallel orientation) is initially
away from the surface with a hight $h_{\rm H_2}$.}
\end{figure}

\begin{figure}
\includegraphics[width=0.8\textwidth]{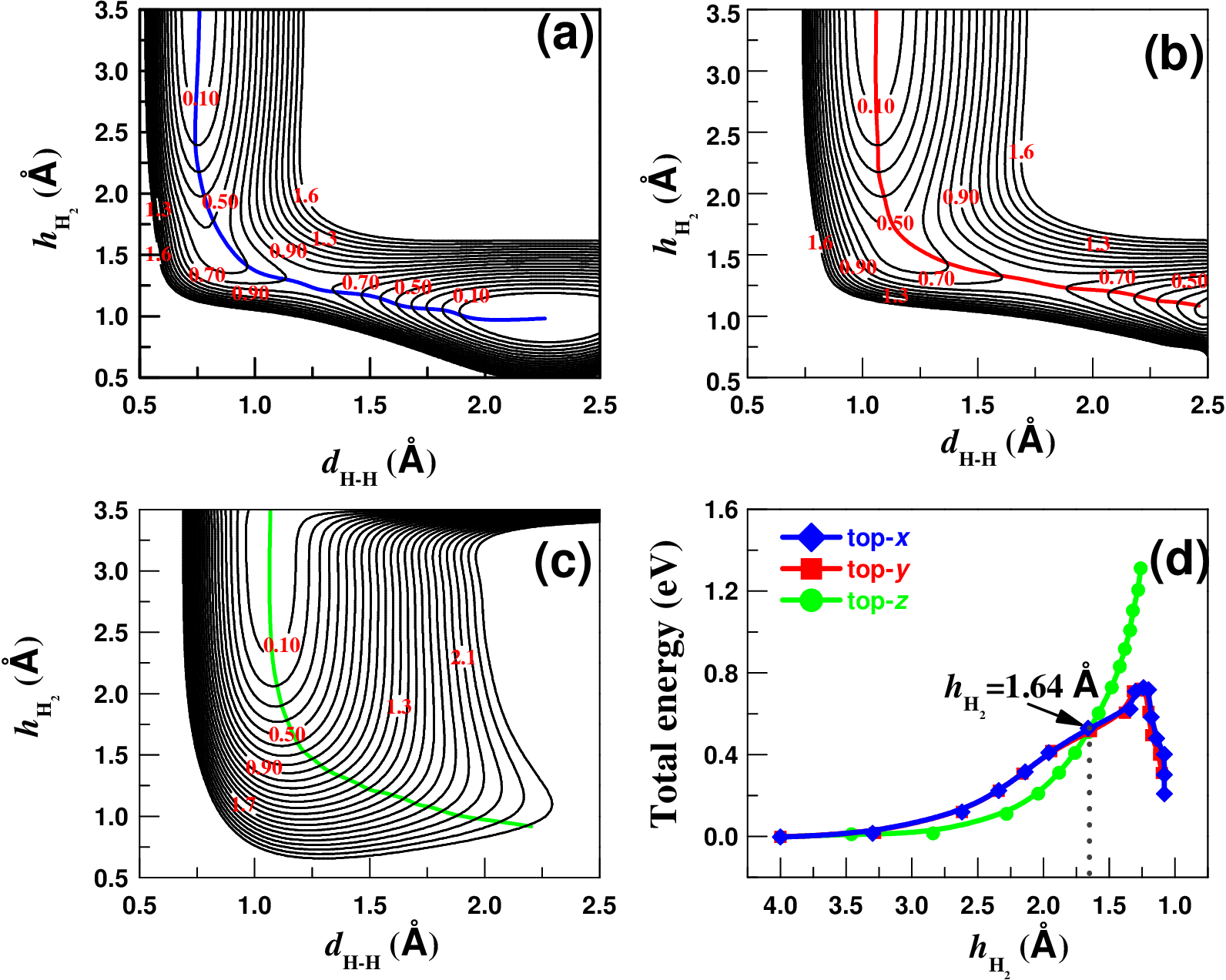}
\caption{(Color online). Contour plots of the two-dimensional cuts
of the potential energy surfaces (PESs) for the H$_{2}$/Be(0001)
system as a function of the bond length ($d_{\rm H-H}$) and
molecular height of H$_2$ ($h_{\rm H_2}$), along the top-$x$ (a),
$y$ (b) and $z$ (c) channels. (d) The corresponding minimum energy
paths for the dissociative adsorption of molecular H$_2$ along the
top-$x,y,z$ channels. In all the figures, the total energy of an
isolated H$_2$ molecule plus a clean Be(0001) surface is set to
zero.}
\end{figure}

After geometry optimization for the Be(0001) surface, we build our
model to study the two-dimensional (2D) PES cuts for H$_{2}$ on the
relaxed Be surface. As shown in Fig. 1, there are four different
high-symmetry sites on the Be (0001) surface, respectively the top,
bridge (bri), hcp and fcc hollow sites. At each surface site, an
adsorbed H$_{2}$ has three different high-symmetry orientations,
respectively along the $x$ (i.e., [$11\bar{2}0$]), $y$ (i.e.,
[$\bar{1}100$]), and $z$ (i.e., [$0001$]) directions. Herein, we use
top-$x,y,z$, bri-$x,y,z$, hcp-$x,y,z$ and fcc-$x,y,z$ respectively
to represent the twelve high-symmetry channels for the adsorption of
H$_{2}$ on the Be surface. We have also tested several low-symmetry
adsorption channels, and found that the dissociation energy barriers
for H$_{2}$ along these low-symmetry channels are always larger than
that along high-symmetry channels. Similar results have also been
obtained for the O$_2$/Pb(111) system where molecular adsorption of
O$_2$ only occurs at surface high-symmetry sites \cite{Yang08}.
Therefore, we will only discuss the PES cuts along the high-symmetry
channels.

From our PES calculations, we find that there are no molecular
adsorption states for H$_2$ on the Be(0001) surface. This finding is
similar to the adsorption of H$_2$ on the Mg(0001) surface
\cite{Vegge2004,Wu2008}. The calculated 2D PES cuts along the
top-$x$ ,$y$ and $z$ channels are respectively listed in Figs. 2(a),
(b) and (c). The other three PES cuts along the bri-, hcp- and
fcc-$z$ channels have similar energy distributions with the top-$z$
channel, and the energies needed to separate the two H atoms to 2.0
\AA~ are all about 2.08 eV for these four channels. The top-$x$ and
$y$ channels are two nearly degenerate channels, as we can see from
Fig. 2(d) that the minimum energy path along these two channels are
the same with different heights. These two channels also have the
lowest energy barriers for the dissociation of H$_2$ among the
twelve adsorption channels. Herein, the dissociation of H$_2$
molecules on the clean Be(0001) surface is different from on the
clean Mg(0001) surface, where the bri-$y$ channel is the most
energetically favored dissociation channel \cite{Vegge2004,Wu2008}.
Besides, it is interesting to find from Fig. 2(d) that the minimum
energy paths along the top-$x$($y$) and $z$ channels have a cross
point at the molecular height of 1.64 \AA, which means that the
H$_2$ molecule would firstly orients perpendicular, and then rotates
to be parallel to the Be surface during its dissociation process.

\begin{figure}
\includegraphics[width=0.8\textwidth]{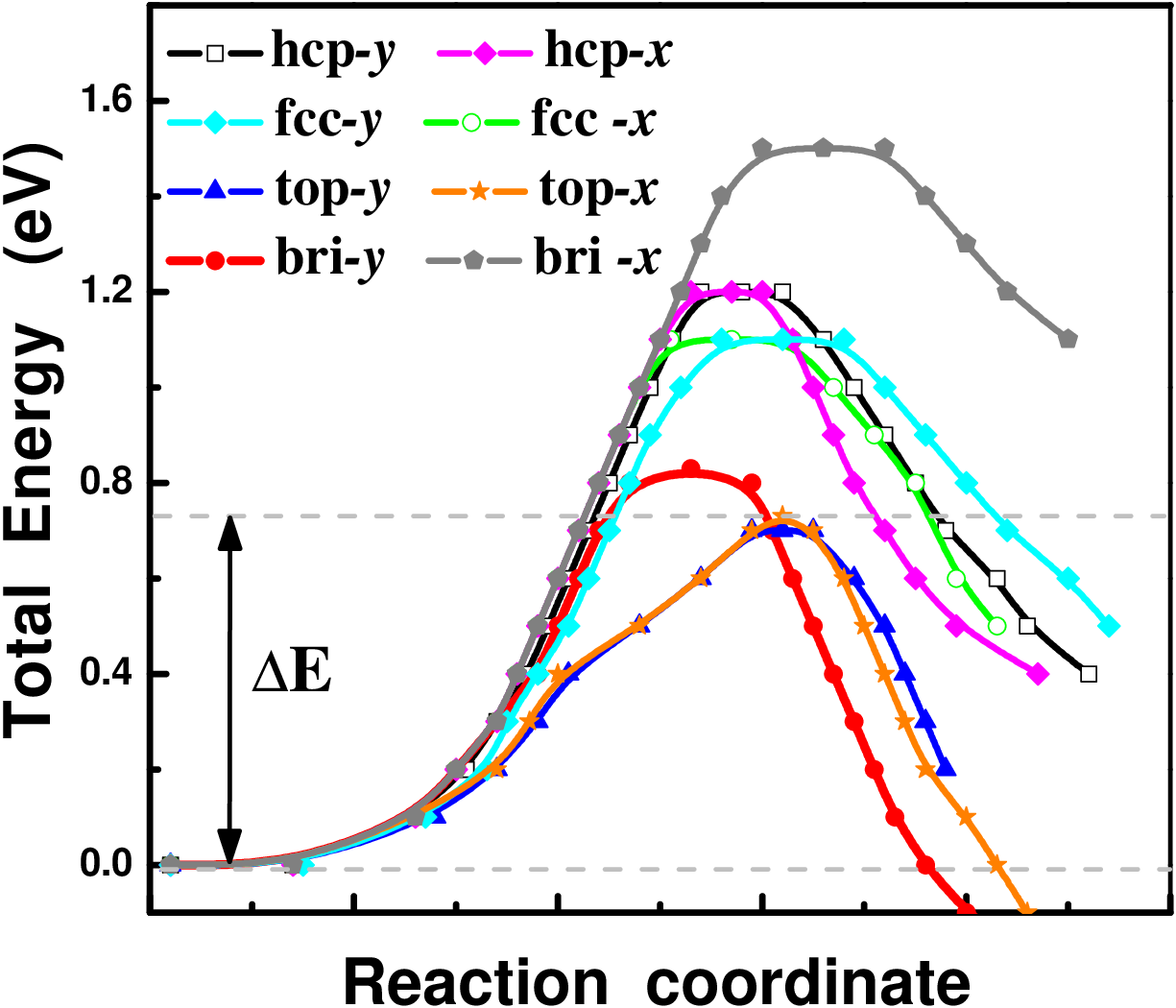}
\caption{(Color online). The minimum energy paths for the
dissociative adsorption of molecular H$_2$ along all the considered
adsorption channels with high symmetries. The total energy of an
isolated H$_2$ molecule plus a clean Be(0001) surface is set to
zero.}
\end{figure}

\begin{table}
\caption{The dissociation energy barriers for $H_2$ along the high
symmetry adsorption channels, and the corresponding bond length
($d_{\rm H-H}$ (TS)) and molecular height of H$_2$ ($h_{\rm H_2}$
(TS)) at the transition states.}
\begin{tabular}{cccc}
  \hline
   adsorption channel & $\Delta$E (eV) & $d_{\rm H-H}$(TS) (\AA)~ & $h_{\rm H_2}$(TS) (\AA)~ \\
  \hline
   top-$x$ & 0.75 & 1.24 & 1.26 \\
   top-$y$ & 0.75 & 1.74 & 1.26 \\
   bri-$x$ & 1.62 & 1.24 & 1.40 \\
   bri-$y$ & 0.91 & 0.88 & 1.40 \\
   hcp-$x$ & 1.28 & 1.04 & 1.28 \\
   hcp-$y$ & 1.29 & 1.04 & 1.32 \\
   fcc-$x$ & 1.18 & 1.06 & 1.26 \\
   fcc-$y$ & 1.22 & 1.06 & 1.26 \\
  \hline
\end{tabular}\label{Ead}
\end{table}

The minimum energy path along the top-, bri-, hcp- and fcc-$x,y$
channels are drawn together in Fig. 3. The corresponding
dissociation energy barrier ($\Delta$E) along these channels are
summarized in Table I. Besides of the top-$x,y$ channels, along
which H$_2$ molecules can easily dissociate at room temperatures in
front of the 0.75 eV energy barrier, the bri-$y$ channel is also a
probable dissociation path since its energy barrier is only 0.16 eV
higher than the top-$x$($y$) channel. Remarkably, the lowest energy
barrier for dissociation of H$_2$ molecules is much smaller on the
Be(0001) surface than on the Mg(0001) surface
\cite{Vegge2004,Wu2008}. This is probably a good news for hydrogen
storage researchers. Form Fig. 3 and Table I, one may also note that
the dissociation energy barriers for H$_2$ is always larger at
surface hcp hollow sites than fcc hollow sites. Considering that
there are more electrons distributing at the surface hcp hollow site
\cite{Wang09}, this finding reflects that charge transfer from the
Be surface to H$_2$ is important for the dissociation of H$_2$. The
bond length ($d_{\rm H-H}$ (TS)) and molecular height of H$_2$
($h_{\rm H_2}$ (TS)) at the corresponding transition states are also
included in Table I. Although the interactions between H atoms and
the Be(0001) surface have long been studied, our results concerning
the dissociation energy barrier and minimum energy paths are still
meaningful.

\section{Diffusion, penetration and distribution of dissociated hydrogen atoms}

The PES cut for one H atom on the Be(0001) surface is also
calculated to study its diffusion characters. Our calculated PES is
shown in Fig. 4(a), with the energy path crossing the four
high-symmetry surface sites shown in Fig. 4(b). One can see that the
hcp and fcc hollow sites are two minimum energy locations for the H
atom, while the bridge site is the saddle point in the diffusion
pathway, and the top sites are maxima of the PES. As shown in Fig.
4(b) the energy barrier for one H atom to diffuse from the hcp (fcc)
to fcc (hcp) hollow site is 0.35 (0.28) eV, which is very small and
can be easily overcome at room temperatures. So after dissociation,
the H atoms will easily diffuse around on the Be(0001) surface.

\begin{figure}
\includegraphics[width=0.8\textwidth]{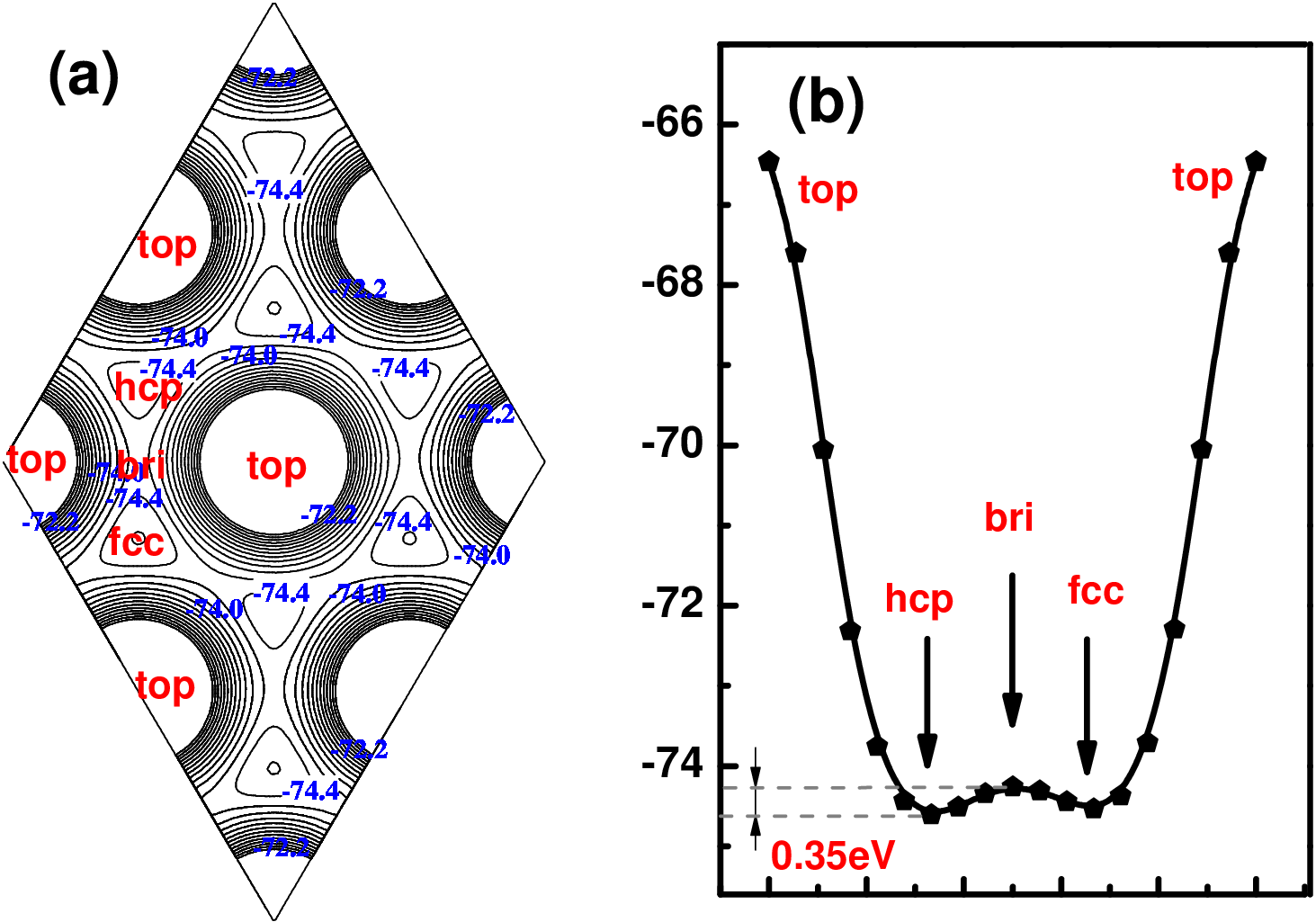}
\caption{(Color online). (a) Contour plots of the potential energy
surface distribution for one H atom on the Be(0001) surface, with
the height of $h_{\rm H_2}$ = 0.85 \AA. (b) The energy curve for one
H atom crossing the four high-symmetry surface sites on the Be(0001)
surface.}
\end{figure}

As has been pointed out, the adsorption of H atoms energetically
prefers the surface hollow sites at low coverage on the Be(0001)
surface \cite{Stumpf1995}. We then study the penetration of H atoms
from these surface hcp and fcc hollow sites to the subsurface sites.
There are three different kinds of high-symmetry subsurface sites.
The octahedral site (octa) lies just underneath the on-surface fcc
site, and one tetrahedral site (tetra) lies below the on-surface hcp
site. A second tetrahedral site (tetra*) is located directly below a
first-layer metal atom. However, the direct hydrogen penetration
into the tetra* site from the on-surface adsorption without
bypassing the other subsurface sites is very unfavorable, since this
site is located beneath a surface Be atom. So the penetration into
the tetra* site will not be discussed. And our calculated results
for penetrations into the other two subsurface sites are shown in
Figs. 5(a) and (b) respectively. It is clearly shown in Fig. 5(a)
that there is no stable adsorption states for atomic H in the
subsurface tetra site, and the energy barrier for one H atom to
penetrate through the Be surface is larger than 1.5 eV. As shown in
Fig. 5(b), for subsurface adsorption, the H atom will stay at the
octa site. However, the energy barrier from on-surface fcc hollow
site to subsurface octa site is as large as 1.25 eV. In addition,
the subsurface adsorption state is much less stable than the
on-surface adsorption state at the fcc hollow site, and the energy
barrier for the H atom to penetrate from subsurface back to
on-surface site is very small ($\sim$ 0.07 eV). So after
dissociation, the H atoms have a much larger probability to stay on
the clean Be(0001) surface than penetrating into the subsurface
sites.

\begin{figure}
\includegraphics[width=0.8\textwidth]{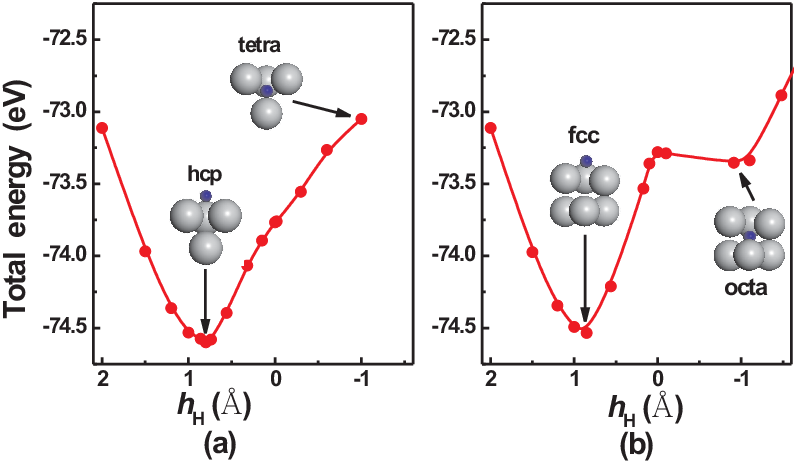}
\caption{(Color online). The energy curve for one H atom penetrating
from the surface hcp to subsurface tetra sites (a), and from the
surface fcc to subsurface octa sites (b).}
\end{figure}

After the systematic studies on the diffusion and penetration energy
barriers, we then analyze the distribution properties for the
dissociated H atoms. Since the penetration of atomic H is found to
be very hard, and the subsurface adsorption is always less stable
than on-surface adsorption, we will mainly discuss the distribution
for on-surface adsorption of H atoms. And we focus on the total
energy calculation for the adsorption system of dissociated H atoms
at different hollow sites. Figures 6(a), (b) and (c) respectively
shows three different adsorption structures after geometry
optimizations, in which the two dissociated H atoms respectively
adsorb at two adjacent fcc and hcp hollow sites, two separate fcc
hollow sites and two separate hcp hollow sites. Obviously, the H
atoms are more clustered to each other in Fig. 6(a) than in Figs.
6(b) and (c). And the structural difference between Figs. 6(b) and
(c) lies in that the H atoms residue at fcc and hcp hollow sites
respectively. Our calculated total energies are respectively -77.70,
-77.98 and -78.07 eV for the three distributions in Figs. 6(a), (b)
and (c). It means that the adsorption state of two H atoms always
has a higher total energy when they are clustered. So after
dissociation, the H atoms will not cluster with each other, instead,
they will distribute uniformly on the clean Be(0001) surface. In
addition, the adsorption of H atoms at surface hcp hollow sites are
always stabler than at surface fcc hollow sites.

\begin{figure}
\includegraphics[width=0.8\textwidth]{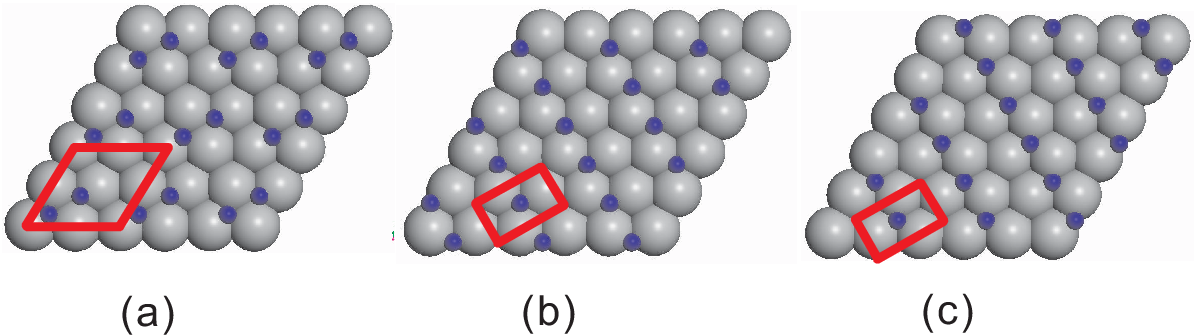}
\caption{(Color online). Three kinds of configurations for atomic H
adsorbates at the same coverage of 0.5 ML. The H adatoms in (a) is
arranged to be clustered, while in (b) and (c) have uniform
distributions. The calculated total energies show that the
adsorption in (c) is the most stable.}
\end{figure}

\section{Adsorption of H$_2$ on hydrogen-preadsorbed Be(0001) surface}

\begin{figure}
\includegraphics[width=0.8\textwidth]{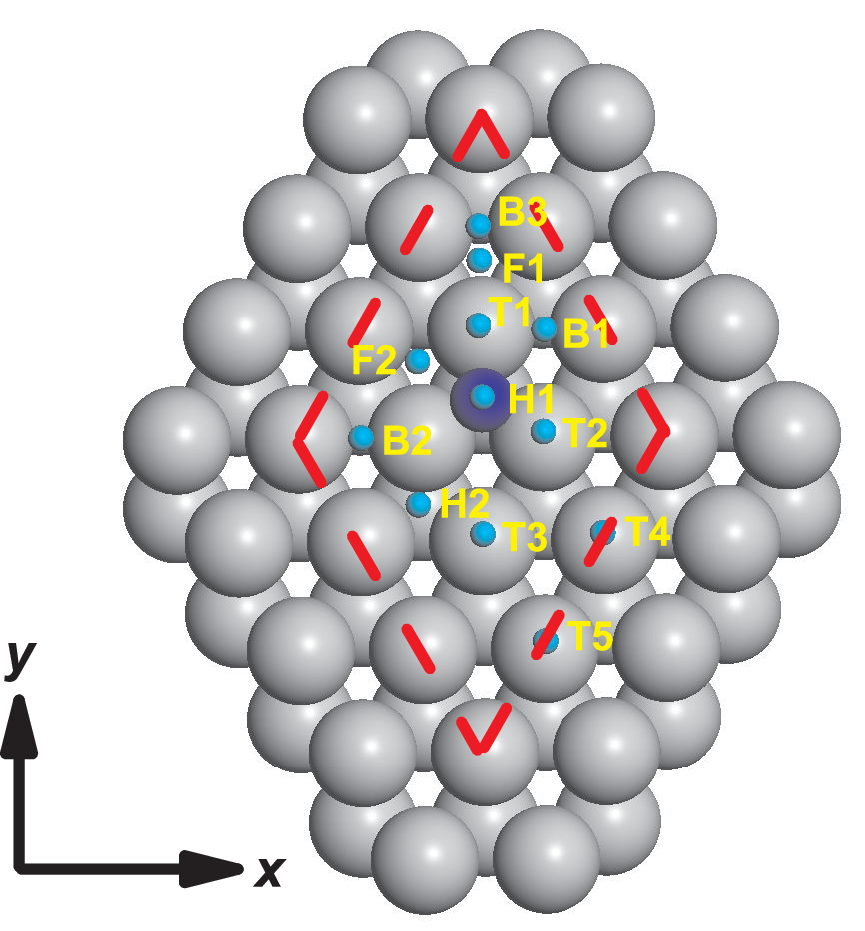}
\caption{(Color online). The schematic model for studying the
adsorption of a H$_2$ molecule on the hydrogen-preadsorbed Be(0001)
surface. The blue and grey balls respectively represent H and Be
atoms. And the light green balls are used to represent the high
symmetry sites, that are denoted as T1$\sim$T5, B1$\sim$B3,
F1$\sim$F2 and H1$\sim$H2.}
\end{figure}

From the previous results, we see that H$_2$ molecules easily
dissociate on the Be(0001) surface, and the dissociated H atoms
tends to distribute uniformly on the surface. Herein, we further
study the adsorption of H$_2$ molecules on the hydrogen-preadsorbed
Be(0001) surface. The preadsorbed H atom is set at a surface hcp
hollow site, since it is the most energetically favored adsorption
site for H atoms at low coverage \cite{Stumpf1995}. And in order to
compare the adsorption of H$_2$ molecules near to and away from the
preadsorbed H atom, a $3\times3$ supercell with 9 Be atoms at each
monolayer is adopted, with the k-points grid for integrations over
the Brillouin zone changed to be $7\times7\times1$. Our structural
model for the hydrogen-preadsorbed Be(0001) surface is shown in Fig.
7, with the considered high symmetry sites respectively denoted as
T1$\sim$T5, B1$\sim$B3, F1$\sim$F2 and H1$\sim$H2. Since a H$_2$
molecule has different possible orientations while adsorbing at each
high symmetry sites, we here use for example, T1-$x$ to represent
for the adsorbing channel of a H$_2$ molecule orienting along the
$x$ (i.e., [$11\bar{2}0$]) direction at the T1 site.

\begin{figure}
\includegraphics[width=0.8\textwidth]{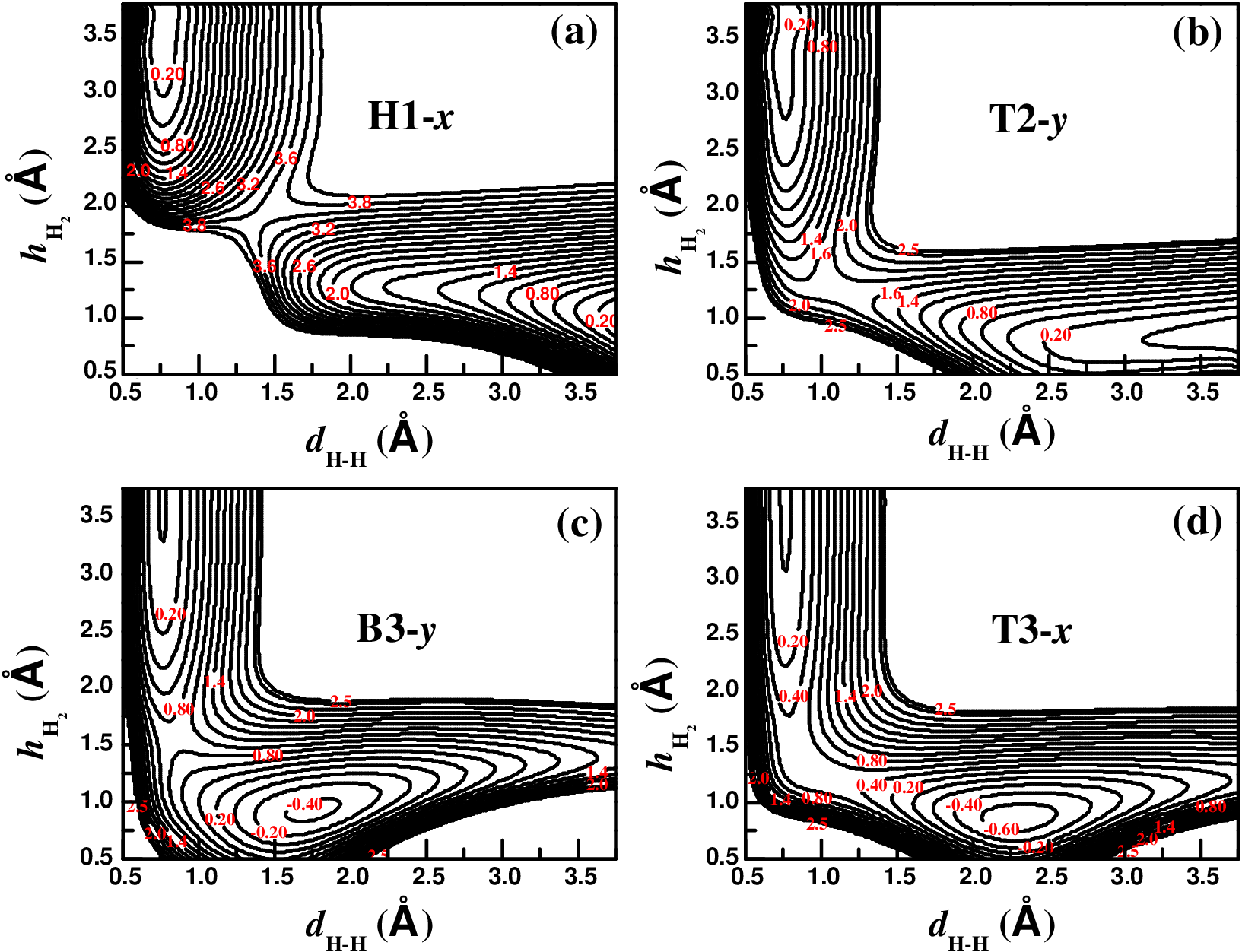}
\caption{(Color online). Contour plots of the two-dimensional cuts
of the potential energy surfaces (PESs) for the H$_{2}$/Be(0001)
system as a function of the bond lengths ($d_{\rm H-H}$) and the
heights ($h_{\rm H_2}$), along the H1-$x$ (a), T2-$y$ (b), B3-$y$
and T3-$x$ (d) channels.}
\end{figure}

Among all the calculations, we find no molecular adsorption states
of H$_2$ on the hydrogen-preadsorbed Be(0001) surface. The
calculated 2D-PES cuts for the H1-$x$, T2-$y$, B3-$y$ and T3-$x$
channels are respectively shown in Figs. 8(a), (b), (c) and (d). As
shown in Fig. 8(a), the dissociation energy barrier for a H$_2$
molecule directly on the precovered H atom is 3.61 eV. Since the
dissociation energy barrier for a H$_2$ molecule at the hcp hollow
site of a clean Be(0001) surface is only 1.28 eV as been listed in
Table I, the preadsorbed H atom actually makes the dissociation of
H$_2$ molecule on top of it much harder. Moreover, as shown in Fig.
8(b), the dissociation energy barrier for a H$_2$ molecule around
the preadsorbed H atom along the T2-$y$ channel is 1.70 eV. In
comparison with the energy barrier of 0.75 eV for a H$_2$ molecule
at the top site of a clean Be(0001) surface, the dissociation of a
H$_2$ molecule becomes harder too around a preadsorbed H atom.
However, when the H$_2$ molecule is enough far away from the
preadsorbed H atom, the dissociation energy barrier can be lower
than that on a clean Be(0001) surface. As shown in Figs. 8(c) and
(d), the dissociation energy barrier for a H$_2$ molecule along the
B3-$y$ and T3-$x$ channels are respectively 0.85 and 0.50 eV, which
are smaller than the dissociation energy barriers of a H$_2$
molecule at the bridge and top site of a clean Be(0001) surface (i.
e., 0.91 and 0.75 eV respectively). The T3-$x$ channel is also the
adsorption channel with the lowest energy barrier for dissociation
of H$_2$ on the preadsorbed Be(0001) surface. Considering that the
dissociation of H$_2$ molecules should not be affected by a
preadsorbed H atom far away from them, and the dissociation energy
barrier of a H$_2$ molecule at the top site of a clean Be(0001)
surface is 0.75 eV, the dissociation energy barrier for a H$_2$
molecule will go back to 0.75 eV when it is very far away from the
preadsorbed hydrogen atom and adsorb at surface top sites.
Therefore, we can see that the horizontal distances from the
preadsorbed hydrogen atoms ($d_0$) have critical effects on the
dissociation energy barrier of adsorbing H$_2$ molecules. The
dissociation energy barriers ($\Delta$E) for an H$_2$ molecule along
all the considered adsorption channels on the hydrogen-preadsorbed
Be(0001) surface are summarized in Table II, together with the bond
length ($d_{\rm H-H}$ (TS)) and molecular height of H$_2$ ($h_{\rm
H_2}$ (TS)) at the corresponding transition states.

\begin{table}
\caption{The dissociation energy barriers for H$_2$ along the high
symmetry adsorption channels on the hydrogen-preadsorbed Be(0001)
surface, geometries for the corresponding transition states and
corresponding horizontal distances from the preadsorbed hydrogen
atom.}
\begin{tabular}{ccccc}
  \hline
   adsorption & \multirow{2}*{$\Delta$E (eV)} & \multirow{2}*{$d_{\rm H-H}$(TS) (\AA)~} & \multirow{2}*{$h_{\rm H_2}$(TS) (\AA)~} & \multirow{2}*{$d_0$
   (\AA)~} \\
   channel &  &  &  & \\
  \hline
   T1-$x$ & 1.54 & 1.14 & 1.36 & 1.31 \\
   T2-$x$ & 0.69 & 1.89 & 1.23 & 1.31 \\
   T2-$y$ & 1.70 & 1.15 & 1.24 & 1.31 \\
   T3-$x$ & 0.50 & 0.95 & 1.32 & 2.62 \\
   T3-$y$ & 0.68 & 1.22 & 1.29 & 2.62 \\
   T4-$x$ & 0.58 & 0.85 & 1.56 & 3.46 \\
   T4-$y$ & 0.60 & 0.88 & 1.56 & 3.46 \\
   T5-$x$ & 0.65 & 1.16 & 1.30 & 4.72 \\
   T5-$y$ & 0.56 & 1.06 & 1.38 & 4.72 \\
   B1-$y$ & 1.21 & 0.90 & 1.45 & 1.73 \\
   B2-$y$ & 0.94 & 0.85 & 1.56 & 2.36 \\
   B3-$y$ & 0.85 & 0.85 & 1.57 & 3.27 \\
   F1-$x$ & 2.06 & 1.34 & 1.36 & 1.31 \\
   F2-$x$ & 1.01 & 0.92 & 1.52 & 2.62 \\
   H1-$x$ & 3.61 & 1.45 & 1.91 & 0.00 \\
   H1-$y$ & 3.58 & 1.43 & 1.91 & 0.00 \\
   H2-$x$ & 1.14 & 1.00 & 1.35 & 2.27 \\
   H2-$y$ & 1.28 & 1.01 & 1.36 & 2.27 \\
  \hline
\end{tabular}\label{Ead}
\end{table}

To illustrate more specifically the dependence of the dissociation
energy barrier for H$_2$ molecules on their horizontal distances
from the preadsorbed hydrogen atom (i.e., $d_0$), we then summarize
the dissociation energy barriers for H$_2$ molecules along all the
considered adsorption channels together and show them in Fig. 9.
Figure 9(a) shows the dissociation energy barriers along all the
top-$x$ (T1$\sim$T5-$x$) and top-$y$ (T1$\sim$T5-$y$) channels, from
which one can see clearly that the dissociation energy barrier has
the largest values when the H$_2$ molecules are directly on top of
the preadsorbed hydrogen atom, and the smallest values when the
H$_2$ molecules are about 3.50 \AA~ far away from the preadsorbed
hydrogen atom. When the H$_2$ molecules are over 4.00 \AA~ far away
from the preadsorbed hydrogen atom, the dissociation energy barriers
for them become larger again and approach the values on the clean
Be(0001) surfaces. In this case, the dependence of the dissociation
energy barriers for H$_2$ molecules on their horizontal distances
from the preadsorbed hydrogen atom is clear. The dissociation energy
barrier for H$_2$ molecules along other adsorption channels are
summarized in Fig. 9(b), from which the lowering down of the
dissociation energy barrier with increasing $d_0$ is also very
clear. However, the increasing of the dissociation energy barrier
with very large $d_0$ is unseen because of the limitation of our
cell size.

\begin{figure}
\includegraphics[width=0.8\textwidth]{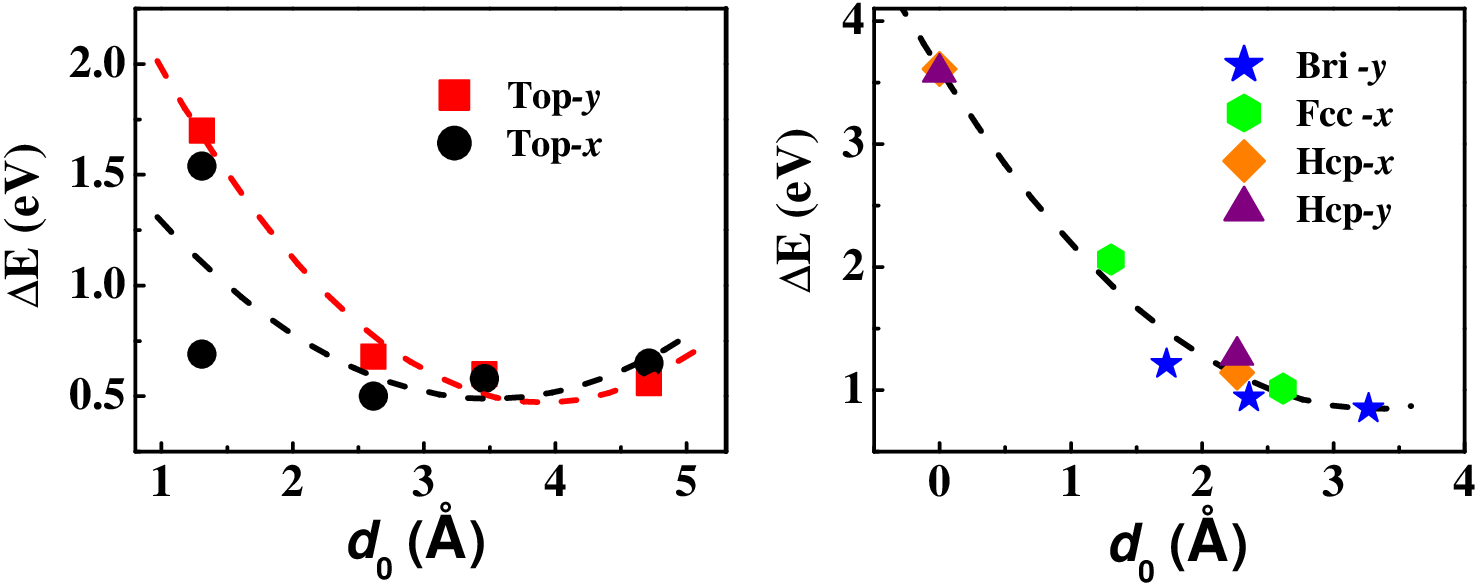}
\caption{(Color online). The dissociation energy barrier for H$_2$
molecules along the T1$\sim$T5-$x,y$ channels (a) and other
considered channels (b) as a function of their horizontal distances
from the preadsorbed hydrogen atom.}
\end{figure}

\section{Conclusions}

In conclusion, we have systematically investigated the dissociation
of H$_2$ molecules on the clean and hydrogen-preadsorbed Be(0001)
surfaces by using the DFT methods. And our paper perfectly describes
the dissociation behaviors for H$_2$ molecules on the Be(0001)
surface.

From the calculated 2D PES cuts for H$_2$ molecules on the clean
Be(0001) surface, we have found that surface top-$x,y$ channels are
the most energetically favorable dissociation channels, and the
lowest dissociation energy barrier is 0.75 eV for the adsorbing
H$_2$ molecules. Besides, the dissociation of H$_2$ along the
bri-$y$ channel is also very probable to happen with the energy
barrier of 0.91 eV. During our calculations, we have also found that
the dissociation energy barrier for H$_2$ at the hcp hollow sites is
always smaller than at the fcc hollow sites on the clean Be(0001)
surface.

For the dissociated hydrogen atoms, we have revealed that the
diffusion energy barrier is as small as 0.35 eV, while the
penetration energy barrier is as large as 1.25 eV, so the
dissociated H atoms can easily diffuse around on the Be(0001)
surface without penetrating into the subsurface sites. We have also
revealed that the adsorption structure with uniform distributions of
dissociated hydrogen atoms always has a lower energy.

Based on these studies, we have then further calculated the 2D PES
cuts for H$_2$ molecules on the hydrogen-preadsorbed Be(0001)
surface, which shows that the dissociation energy barriers for H$_2$
molecules depend critically on their horizontal distances from the
preadsorbed hydrogen atom. The dissociation of H$_2$ molecules
directly on the preadsorbed hydrogen atom is very hard, but the
lowest dissociation energy barrier for H$_2$ on the
hydrogen-preadsorbed Be surface is 0.25 eV smaller than that on a
clean Be surface.

\begin{acknowledgments}
This work was supported by the NSFC under Grant No. 10604010, No.
60776063, No. 50471070, No. 50644041, and by a fund from CAEP.
\end{acknowledgments}

\end{document}